\documentclass[onecolumn]{aastex63}
\usepackage{amsmath,amstext}
\usepackage[T1]{fontenc}
\usepackage{apjfonts} 
\usepackage[figure,figure*]{hypcap}
\usepackage{CJKutf8}

\submitjournal{The Planetary Science Journal (AAS)}

\shortauthors{Pike et al.}

\begin{document}

\title{A Deep Search for Binary TNOs}

\author[0000-0003-4797-5262]{Rosemary~E. Pike}
\affiliation{Institute of Astronomy and Astrophysics, Academia Sinica; 11F of AS/NTU Astronomy-Mathematics Building, No. 1 Roosevelt Rd., Sec. 4, Taipei 10617, Taiwan}
%\affiliation{Harvard \& Smithsonian Center for Astrophysics; 60 Garden Street, Cambridge, MA, 02138, USA}
\affiliation{Smithsonian Astrophysical Observatory; 60 Garden Street, Cambridge, MA, 02138, USA}
\author[0000-0003-0386-2178]{Jayatee Kanwar}
\affiliation{Institute of Astronomy and Astrophysics, Academia Sinica; 11F of AS/NTU Astronomy-Mathematics Building, No. 1 Roosevelt Rd., Sec. 4, Taipei 10617, Taiwan}
\author[0000-0003-4143-8589]{Mike Alexandersen}
\affiliation{Institute of Astronomy and Astrophysics, Academia Sinica; 11F of AS/NTU Astronomy-Mathematics Building, No. 1 Roosevelt Rd., Sec. 4, Taipei 10617, Taiwan}
%\affiliation{Harvard \& Smithsonian Center for Astrophysics; 60 Garden Street, Cambridge, MA, 02138, USA}
\affiliation{Smithsonian Astrophysical Observatory; 60 Garden Street, Cambridge, MA, 02138, USA}
\author[0000-0001-7244-6069]{Ying-Tung Chen (\begin{CJK*}{UTF8}{bkai}陳英同\end{CJK*})}
\affiliation{Institute of Astronomy and Astrophysics, Academia Sinica; 11F of AS/NTU Astronomy-Mathematics Building, No. 1 Roosevelt Rd., Sec. 4, Taipei 10617, Taiwan}
\author[0000-0003-4365-1455]{Megan E. Schwamb}
\affiliation{Astrophysics Research Centre, School of Mathematics and Physics, Queen's University Belfast, Belfast BT7 1NN, UK}
\affiliation{Institute of Astronomy and Astrophysics, Academia Sinica; 11F of AS/NTU Astronomy-Mathematics Building, No. 1 Roosevelt Rd., Sec. 4, Taipei 10617, Taiwan}
\collaboration{5}{}

\begin{abstract}

The Latitude Density Search utilized Hyper Suprime-Cam on Subaru Telescope to discover 60 moving objects in the outer Solar System, 54 of which have semi-major axes beyond 30 AU. 
The images were acquired in exceptional seeing (0.4$\arcsec$) and reached a detection limit of $m_r\simeq25.2$.
The two night arcs were used to calculate orbits which are poorly constrained, however, the distance and inclination are the parameters best constrained by short arcs and a reasonable determination can be made of which objects are cold classical TNOs and which are dynamically excited.
We identify 10 objects as likely cold classical objects.
We searched all of the detections for binary sources using a trailed Point Spread Function subtraction method, and identified one binary object with a separation of 0.34$\arcsec$ and a secondary with 17\% the brightness of the primary (2.0 magnitudes fainter).
This is the brightest TNO in the sample, the previously known object 471165 (2010 HE$_{79}$), which has a dynamically excited orbit.
Because of the excellent seeing, this search was sensitive to binaries with 0.34$\arcsec$ separation and a brightness of $\ge$50\% the primary brightness for 7 objects, including one cold classical.
This gives an intrinsic binary fraction of $\sim$17\% (1 of 6) for the dynamically excited objects within our detection limits.
The trailed point spread function subtraction method to identify binaries, fit the two components, and determine the sensitivity limits, used in the Latitude Density Search is a useful tool that could be more broadly applied to identify binary TNOs and track known binary TNO orbits.

\end{abstract}

\section{Introduction}

Trans-Neptunian objects (TNOs) populate the distant icy solar system beyond Neptune, and the characteristics of these small icy bodies provide useful insights into the formation of planetesimals in the Solar System.
TNOs include both a sub-population with low inclination ($i$) and low eccentricity ($e$), referred to as cold classical TNOs, and several sub-populations with higher inclinations and/or eccentricities, which are referred to as hot or dynamically excited TNOs \citep{gladman2008}.
The cold classical TNOs have distinct orbits \citep{brown2001} and are thought to have formed in their current locations \citep[e.g.][]{levisonStern2001}.
These cold classicals have experienced only slight stirring during Neptune's migration \citep{nesvorny2015b} and minimal collisional grinding \citep{nesvorny_etal2011}.
The dynamically excited population includes hot classical TNOs (with similar semi-major axes, $a$, to the cold classicals), as well as resonant, scattering, and detached TNOs \citep{gladman2008}.
All of the dynamically excited objects are likely implanted into the Kuiper belt from an inner disk, through interactions with Neptune, which increased their inclinations and eccentricities \citep[e.g.][]{levisonmorbidelli2008}.

In addition to different orbital characteristics, the dynamically excited TNOs and cold classical TNOs have distinct physical characteristics.
Cold classical objects typically have a higher albedo ($\sim14$\%), while dynamically excited TNOs have a broader range of albedos with a lower average albedo \citep[$\sim8.5$\%,][]{vilenius2018}.
Cold classical objects also have distinct surface colors in the optical to near-infrared, which likely implies a difference in surface composition between the populations \citep[e.g.][]{tegler2000,pike2017z,colossos, marsset2019}.
The cold classicals have also been found to harbor a large fraction of binary objects, approximately 30\% compared to 3--10\% for dynamically excited objects, although many of these are close binaries difficult to detect from ground-based surveys \citep{noll2008}.
Close encounters with Neptune would have disrupted the wide binary pairs found in the cold classical population \citep{parker2010}.
The difference in binary fraction between the cold classicals and dynamically excited objects is likely linked to their different formation and evolutionary histories \citep{noll2020}.

Previous surveys have found TNOs in wide binaries, where the two components can be optically resolved.
Previous binary searches which targeted a mix of dynamical classes have reported binary detection rates of 1.4\%--11\% depending on the effective resolution of the search, which was typically driven by seeing and plate scale \citep{noll2003, kern2006, kernelliot2006}.
When the components can be resolved, color measurements of binary TNOs have shown that the primary and secondary object have the same colors in the optical and near-infrared \citep{noll2008ssbn,benecchi,marsset2020}.
\citet{benecchi} suggests a co-formation of the binary pair, and not a later capture, as capture models have difficulty preserving the correlated surface colors.
One possible explanation for the high rate of binary formation is that the planetesimals in the Kuiper belt formed via the streaming instability \citep[e.g.][]{youdin2005, johansen2007, chiang2010, nesvorny_youdin_richardson_2010,  nesvorny_etal2019}.
The survival of binary pairs which formed in the outer solar system depends on the lifetime of the disk before planetary migration (and thus the degree of collisional grinding), and the specifics of Neptune's migration \citep{nesvorny_vokrouhlicky2019}.
Understanding how the binary fraction of TNOs of different orbital classes varies with size can provide useful insight into how these objects formed and evolved. 

In this deep two night search, the Latitude Density Search discovered 54 distant solar system objects $5\lesssim H_r \lesssim10$.
We searched the discovery data for binary TNOs.
The survey description and details about the discoveries are presented in Section \ref{sec:discovery}.
In Section \ref{sec:binary_id}, the methods for identifying binary TNOs and fitting the two components of the binary object are described.
The sensitivity of the binary search is explored in Section \ref{sec:sensitivity}, and the results are discussed in Section \ref{sec:discussion}.

\section{Discovery Search}
\label{sec:discovery}

\begin{figure}[h]
\begin{center}
\includegraphics[width=.7\textwidth]{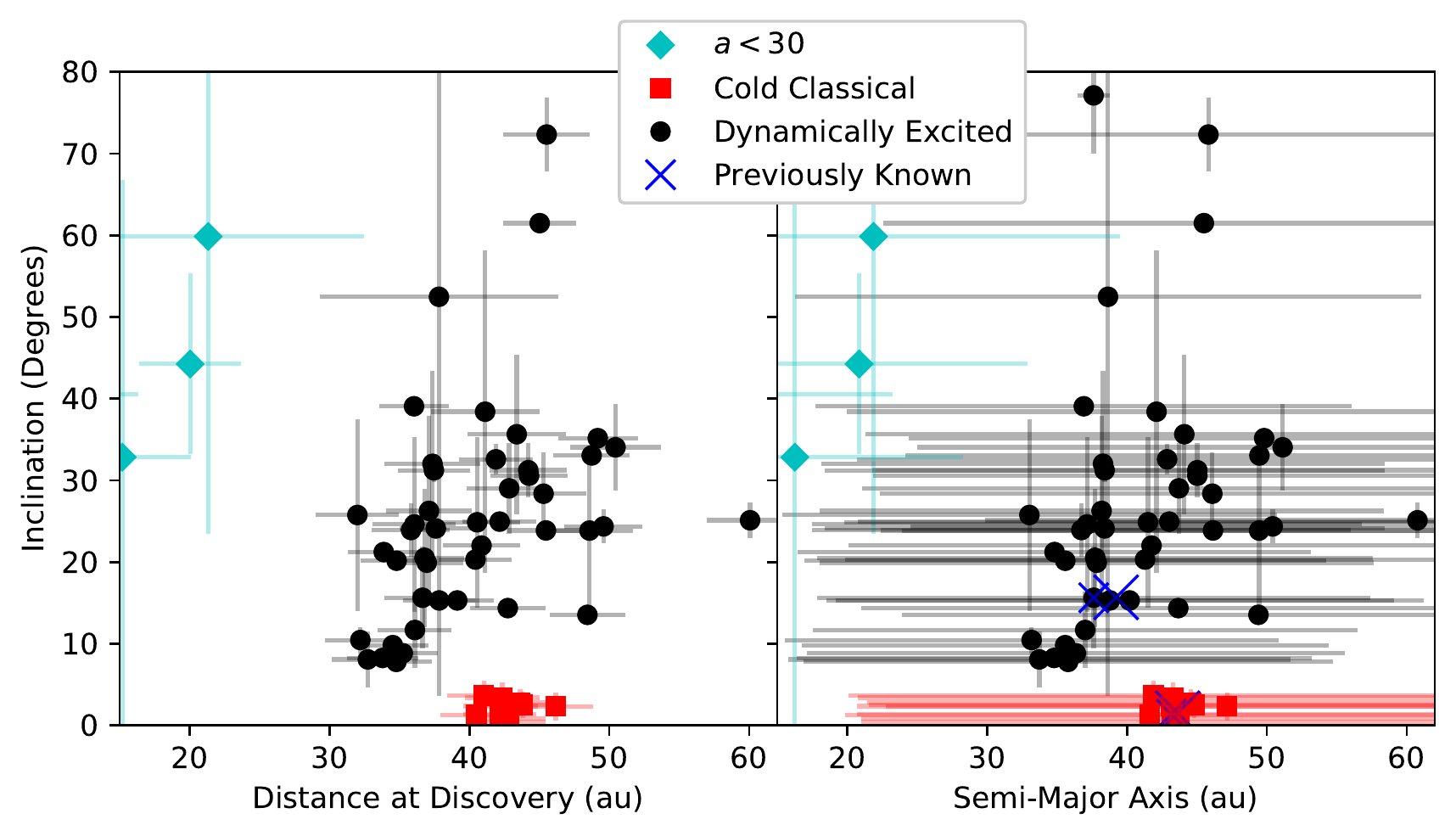}
\caption{
Latitude Density Search discoveries.  \textbf{Left:} The parameters best constrained using a 2 night arc are the orbital inclination and distance at discovery.  The objects we classify as `likely cold classicals' are indicated in red.  Dynamically excited objects (black circles) and Centaurs (cyan diamonds) have higher inclination orbits. \textbf{Right:} The eccentricity is not well measured in a 2 night arc, so the semi-major axis is also poorly constrained. The orbital fits almost exclusively assume a low-$e$ orbit 
(see Table \ref{tnolist}), so for objects where this is the case the calculated semi-major axis is typically quite close to the real value.  The two sets of small and large blue `X's indicate the orbit fit from the 2 night arc and the published orbital parameters for the known objects, respectively, and are well within the formal uncertainties.
}
\label{fig_dai}
\end{center}
\end{figure}

The Latitude Density Search measured the density of faint and distant TNOs at different latitudes off the ecliptic plane and provides an excellent sample which can be compared with TNO distribution models.
The search design, detection pipeline, and results are discussed in more detail in the Latitude Density Search paper \citep{chen2020}; a short summary is provided here.
The observations were executed on June 9-10, 2016 with the 8.2~m Subaru telescope on Maunakea, using the $r$-band filter on Hyper Suprime-Cam (HSC), which has a pixel scale of 0.17$\arcsec$ pixel$^{-1}$ and 104 chips covering a 1.5$^{\circ}$ field of view \citep{HSCpaperCite}.
Images were acquired of 75 fields, with coordinates ranging from centered on the ecliptic plane to 80 degrees off ecliptic.
The images were processed using the HSC pipeline \citep{HSCpipeline, ivezic2019, axelrod2010, juric2017}.
Because the off-ecliptic density of TNOs is significantly lower, more fields were placed at high latitudes in order to ensure the discovery of TNOs at a range of latitudes.
Only one field was placed directly on the ecliptic plane.
Each field image was 120 seconds and was repeated twice per night for two nights.
The moving object search and identification was done in a similar manner to \citet{chen2018}, and required detection in all 4 images.
The search detection limit is $m_r\simeq$25.2 for 5$\sigma$ detections.
The search identified 60 outer Solar System objects, 56 with semi-major axes beyond 30 AU, listed in Table \ref{tnolist} and shown in Figure \ref{fig_dai}.

\begin{table}[h!]
\setlength{\tabcolsep}{2.5pt}
\tiny
\caption{High Latitude Search Discoveries}
\label{tnolist}
\begin{center}
\begin{tabular}{ l l | c  c  c  c  c  c  c c c c c} 
Search Name & MPC Name & a$\pm$$\Delta$ a & e$\pm$$\Delta$ e & i$\pm$$\Delta$ i & d$\pm\Delta$ d & Rate & Trailing & FWHM & $H_r$ & $m_r$& Hill Radius & Likely Classification\\
 &  & (au) &  & (deg) & (au) & $\arcsec$ hr$^{-1}$& $\arcsec$ exp.$^{-1}$ & $\arcsec$ & (mag) & (mag)  & $\arcsec$ on sky & \\ \hline
 F06\_cand00002 & & 20.9$\pm$12.0 & 0.05$\pm$0.67 & 44.2$\pm$11.1 & 20.0$\pm$3.6 & 6.45 & 0.21 & 0.59, 0.60, 0.61, 0.64 & 10.8$^{+0.9}_{-0.7}$ & 23.70$\pm$0.03 & 3.4 & $a<30$\\
F17\_cand00002 & & 43.7$\pm$22.6 & 0.02$\pm$0.57 & 29.0$\pm$5.6 & 42.8$\pm$3.1 & 2.88 & 0.10 & 0.67, 0.59, 0.76, 0.62 & 8.3$^{+0.3}_{-0.3}$ & 24.60$\pm$0.06 & 10.1 & DE\\
F17\_cand00004 & & 43.0$\pm$22.3 & 0.02$\pm$0.57 & 24.9$\pm$0.4 & 42.2$\pm$2.6 & 2.85 & 0.09 & 0.66, 0.62, 0.72, 0.65 & 8.6$^{+0.3}_{-0.3}$ & 24.78$\pm$0.07 & 9.0 & DE\\
F17\_cand00005 & & 46.1$\pm$23.8 & 0.02$\pm$0.57 & 28.4$\pm$5.0 & 45.3$\pm$3.1 & 2.72 & 0.09 & 0.63, 0.60, 0.67, 0.60 & 7.9$^{+0.3}_{-0.3}$ & 24.44$\pm$0.05 & 12.1 & DE\\
F17\_cand00008 & & 45.0$\pm$23.2 & 0.02$\pm$0.57 & 31.2$\pm$3.3 & 44.2$\pm$2.8 & 2.77 & 0.09 & 0.65, 0.58, 0.79, 0.56 & 7.4$^{+0.3}_{-0.3}$ & 23.85$\pm$0.03 & 15.1 & DE\\
F17\_cand00009 & & 46.2$\pm$23.7 & 0.02$\pm$0.57 & 23.9$\pm$0.8 & 45.5$\pm$2.8 & 2.67 & 0.09 & 0.60, 0.59, 0.71, 0.58 & 8.0$^{+0.3}_{-0.3}$ & 24.49$\pm$0.06 & 11.9 & DE\\
F17\_cand00011 & & 38.4$\pm$20.0 & 0.02$\pm$0.57 & 24.1$\pm$0.1 & 37.6$\pm$2.6 & 3.19 & 0.11 & 0.64, 0.59, 0.73, 0.61 & 8.8$^{+0.3}_{-0.3}$ & 24.51$\pm$0.05 & 8.1 & DE\\
F17\_cand00013 & & 60.8$\pm$30.9 & 0.01$\pm$0.56 & 25.1$\pm$2.2 & 60.1$\pm$3.1 & 2.03 & 0.07 & 0.71, 0.76, 0.77, 0.79 & 6.2$^{+0.2}_{-0.2}$ & 23.99$\pm$0.04 & 26.0 & DE\\
F18\_cand00001 & & 44.1$\pm$22.8 & 0.02$\pm$0.58 & 35.6$\pm$9.8 & 43.4$\pm$3.5 & 2.86 & 0.10 & 0.62, 0.62, 0.63, 0.69 & 8.9$^{+0.4}_{-0.3}$ & 25.23$\pm$0.12 & 7.7 & DE\\
F18\_cand00002 & & 13.0$\pm$10.3 & 0.08$\pm$1.02 & 40.5$\pm$19.1 & 12.0$\pm$4.4 & 9.30 & 0.31 & 0.71, 0.68, 0.72, 0.75 & 13.7$^{+2.1}_{-1.4}$ & 24.34$\pm$0.04 & 0.9 & $a<30$\\
F18\_cand00014 & & 38.6$\pm$22.4 & 0.03$\pm$0.72 & 52.5$\pm$48.8 & 37.8$\pm$8.5 & 3.52 & 0.12 & 0.66, 0.63, 0.71, 0.74 & 8.3$^{+1.1}_{-0.9}$ & 24.05$\pm$0.04 & 10.1 & DE\\
F18\_cand00017 & & 42.9$\pm$22.2 & 0.02$\pm$0.57 & 32.6$\pm$1.9 & 41.9$\pm$2.7 & 2.94 & 0.10 & 0.59, 0.61, 0.69, 0.68 & 8.4$^{+0.3}_{-0.3}$ & 24.58$\pm$0.07 & 9.7 & DE\\
F18\_cand00019 & & 38.4$\pm$20.0 & 0.03$\pm$0.57 & 31.2$\pm$1.4 & 37.5$\pm$2.6 & 3.23 & 0.11 & 0.67, 0.65, 0.65, 0.72 & 9.3$^{+0.3}_{-0.3}$ & 24.97$\pm$0.09 & 6.5 & DE\\
F19\_cand00002 & & 45.0$\pm$23.2 & 0.02$\pm$0.57 & 30.6$\pm$0.7 & 44.3$\pm$2.8 & 2.78 & 0.09 & 0.62, 0.58, 0.65, 0.61 & 8.8$^{+0.3}_{-0.3}$ & 25.23$\pm$0.09 & 8.0 & DE\\
F19\_cand00003 & & 51.1$\pm$26.1 & 0.02$\pm$0.57 & 34.0$\pm$5.3 & 50.4$\pm$3.3 & 2.49 & 0.08 & 0.66, 0.58, 0.68, 0.64 & 8.0$^{+0.3}_{-0.3}$ & 25.03$\pm$0.09 & 11.3 & DE\\
F20\_cand00002 & & 49.5$\pm$25.3 & 0.02$\pm$0.56 & 33.0$\pm$0.3 & 48.7$\pm$2.7 & 2.60 & 0.09 & 0.62, 0.64, 0.63, 0.66 & 8.5$^{+0.3}_{-0.2}$ & 25.36$\pm$0.11 & 9.1 & DE\\
F21\_cand00012 & & 49.8$\pm$25.4 & 0.01$\pm$0.56 & 35.2$\pm$0.6 & 49.2$\pm$2.9 & 2.58 & 0.09 & 0.67, 0.65, 0.60, 0.70 & 8.2$^{+0.3}_{-0.2}$ & 25.09$\pm$0.09 & 10.5 & DE\\
F23\_cand00001 & & 36.9$\pm$19.2 & 0.02$\pm$0.57 & 39.1$\pm$0.3 & 36.0$\pm$2.5 & 3.51 & 0.12 & 0.66, 0.63, 0.61, 0.68 & 9.5$^{+0.3}_{-0.3}$ & 24.97$\pm$0.09 & 6.0 & DE\\
F33\_cand00001 & & 21.9$\pm$17.6 & 0.04$\pm$0.87 & 59.8$\pm$36.4 & 21.3$\pm$11.1 & 7.30 & 0.24 & 0.66, 0.64, 0.72, 0.68 & 10.9$^{+3.3}_{-1.9}$ & 24.08$\pm$0.04 & 3.2 & $a<30$\\
F40\_cand00007 & & 44.8$\pm$23.3 & 0.02$\pm$0.57 & 2.5$\pm$1.6 & 43.8$\pm$2.7 & 2.93 & 0.10 & 0.49, 0.51, 0.54, 0.50 & 8.0$^{+0.3}_{-0.3}$ & 24.35$\pm$0.07 & 9.2 & Cold Classical\\
F40\_cand00008 & & 43.7$\pm$22.7 & 0.02$\pm$0.57 & 0.9$\pm$1.3 & 42.8$\pm$2.6 & 3.01 & 0.10 & 0.51, 0.51, 0.56, 0.52 & 8.3$^{+0.3}_{-0.3}$ & 24.59$\pm$0.08 & 7.9 & Cold Classical\\
F40\_cand00009 & 469610 (2004 HF$_{79}$) & 43.2$\pm$22.5 & 0.02$\pm$0.57 & 1.4$\pm$1.3 & 42.2$\pm$2.6 & 3.08 & 0.10 & 0.74, 0.73, 0.78, 0.73 & 6.4$^{+0.3}_{-0.3}$ & 22.63$\pm$0.02 & 18.9 & Cold Classical\\
F40\_cand00014 & & 37.7$\pm$19.9 & 0.03$\pm$0.58 & 20.5$\pm$8.5 & 36.8$\pm$2.8 & 3.49 & 0.12 & 0.47, 0.48, 0.50, 0.50 & 9.0$^{+0.4}_{-0.3}$ & 24.62$\pm$0.07 & 7.4 & DE\\
F40\_cand00017 & & 43.3$\pm$22.5 & 0.02$\pm$0.57 & 3.4$\pm$1.8 & 42.4$\pm$2.7 & 3.02 & 0.10 & 0.69, 0.71, 0.73, 0.70 & 8.7$^{+0.3}_{-0.3}$ & 24.88$\pm$0.13 & 6.7 & Cold Classical\\
F40\_cand00019 & & 38.2$\pm$20.2 & 0.03$\pm$0.59 & 26.2$\pm$11.6 & 37.1$\pm$3.1 & 3.55 & 0.12 & 0.47, 0.47, 0.50, 0.49 & 8.6$^{+0.4}_{-0.3}$ & 24.25$\pm$0.05 & 9.0 & DE\\
F40\_cand00020 & & 41.6$\pm$21.8 & 0.03$\pm$0.57 & 1.3$\pm$1.2 & 40.5$\pm$2.6 & 3.20 & 0.11 & 0.63, 0.53, --, -- & 8.8$^{+0.3}_{-0.3}$ & 24.78$\pm$0.09 & 6.5 & Cold Classical\\
F40\_cand00024 & & 33.7$\pm$17.9 & 0.03$\pm$0.59 & 8.1$\pm$3.4 & 32.7$\pm$2.6 & 3.86 & 0.13 & 0.46, 0.49, 0.50, 0.50 & 9.6$^{+0.4}_{-0.3}$ & 24.67$\pm$0.08 & 5.7 & DE\\
F40\_cand00029 & & 41.9$\pm$21.8 & 0.02$\pm$0.57 & 3.7$\pm$1.8 & 41.0$\pm$2.6 & 3.14 & 0.10 & 0.74, 0.73, 0.76, 0.74 & 8.9$^{+0.3}_{-0.3}$ & 24.96$\pm$0.13 & 6.1 & Cold Classical\\
F40\_cand00030 & & 43.2$\pm$22.4 & 0.02$\pm$0.57 & 2.4$\pm$1.3 & 42.2$\pm$2.6 & 3.00 & 0.10 & 0.51, 0.51, 0.55, 0.52 & 8.3$^{+0.3}_{-0.3}$ & 24.54$\pm$0.08 & 7.8 & Cold Classical\\
F40\_cand00038 & & 43.8$\pm$22.7 & 0.02$\pm$0.57 & 0.4$\pm$0.9 & 42.8$\pm$2.6 & 3.01 & 0.10 & 0.47, 0.47, 0.51, 0.48 & 8.2$^{+0.3}_{-0.3}$ & 24.51$\pm$0.07 & 8.2 & Cold Classical\\
F40\_cand00039 & & 33.0$\pm$17.7 & 0.03$\pm$0.60 & 25.7$\pm$11.8 & 32.0$\pm$3.0 & 4.08 & 0.14 & 0.74, 0.73, 0.78, 0.73 & 8.7$^{+0.4}_{-0.4}$ & 23.64$\pm$0.04 & 8.8 & DE\\
F40\_cand00042 & & 49.5$\pm$25.5 & 0.02$\pm$0.57 & 23.8$\pm$9.7 & 48.6$\pm$3.1 & 2.73 & 0.09 & 0.48, 0.47, 0.50, 0.49 & 6.9$^{+0.3}_{-0.3}$ & 23.75$\pm$0.03 & 19.1 & DE\\
F40\_cand00044 & & 44.6$\pm$23.1 & 0.02$\pm$0.57 & 2.8$\pm$1.6 & 43.6$\pm$2.6 & 2.95 & 0.10 & 0.55, 0.50, 0.55, 0.51 & 8.4$^{+0.3}_{-0.3}$ & 24.74$\pm$0.09 & 7.6 & Cold Classical\\
F40\_cand00048 & & 37.0$\pm$19.5 & 0.03$\pm$0.58 & 11.7$\pm$4.7 & 36.1$\pm$2.6 & 3.51 & 0.12 & 0.51, 0.49, 0.53, -- & 9.7$^{+0.3}_{-0.3}$ & 25.19$\pm$0.13 & 5.5 & DE\\
F40\_cand00052 & & 47.2$\pm$24.4 & 0.02$\pm$0.56 & 2.3$\pm$1.7 & 46.2$\pm$2.7 & 2.79 & 0.09 & 0.59, 0.52, 0.60, 0.53 & 8.4$^{+0.3}_{-0.2}$ & 24.96$\pm$0.12 & 7.7 & Cold Classical\\
F40\_cand00053 & 471165 (2010 HE$_{79})$ & 37.6$\pm$19.8 & 0.03$\pm$0.58 & 15.6$\pm$6.2 & 36.7$\pm$2.7 & 3.48 & 0.12 & 0.50, 0.49, 0.55, 0.50 & 5.3$^{+0.3}_{-0.3}$ & 20.88$\pm$0.00 & 41.2 & DE\\
F40\_cand00056 & & 37.1$\pm$19.6 & 0.03$\pm$0.59 & 24.6$\pm$10.7 & 36.1$\pm$3.0 & 3.63 & 0.12 & 0.52, 0.59, 0.58, 0.59 & 9.6$^{+0.4}_{-0.3}$ & 25.16$\pm$0.15 & 5.6 & DE\\
F40\_cand00057 & & 42.1$\pm$22.1 & 0.02$\pm$0.60 & 38.4$\pm$19.8 & 41.1$\pm$3.9 & 3.30 & 0.11 & 0.56, 0.49, 0.55, 0.51 & 7.5$^{+0.4}_{-0.4}$ & 23.64$\pm$0.03 & 14.5 & DE\\
F42\_cand00006 & & 36.3$\pm$19.2 & 0.03$\pm$0.58 & 8.8$\pm$0.3 & 35.2$\pm$2.5 & 3.46 & 0.12 & 0.55, 0.51, 0.52, 0.53 & 9.7$^{+0.3}_{-0.3}$ & 25.08$\pm$0.13 & 5.5 & DE\\
F42\_cand00009 & & 33.2$\pm$17.6 & 0.03$\pm$0.59 & 10.4$\pm$1.6 & 32.2$\pm$2.5 & 3.79 & 0.13 & 0.53, 0.58, 0.53, 0.54 & 10.0$^{+0.4}_{-0.3}$ & 25.01$\pm$0.11 & 4.8 & DE\\
F42\_cand00021 & & 35.6$\pm$18.8 & 0.03$\pm$0.58 & 9.8$\pm$0.6 & 34.5$\pm$2.5 & 3.53 & 0.12 & 0.56, 0.51, 0.52, 0.54 & 9.9$^{+0.3}_{-0.3}$ & 25.20$\pm$0.15 & 5.0 & DE\\
F42\_cand00031 & & 34.8$\pm$18.4 & 0.03$\pm$0.58 & 8.3$\pm$0.2 & 33.8$\pm$2.5 & 3.61 & 0.12 & 0.48, 0.50, 0.49, 0.48 & 10.0$^{+0.3}_{-0.3}$ & 25.18$\pm$0.13 & 4.8 & DE\\
F42\_cand00042 & & 35.8$\pm$18.9 & 0.03$\pm$0.58 & 7.8$\pm$0.5 & 34.8$\pm$2.5 & 3.53 & 0.12 & 0.47, 0.49, 0.46, 0.48 & 9.8$^{+0.3}_{-0.3}$ & 25.16$\pm$0.12 & 5.2 & DE\\
F42\_cand00061 & & 41.5$\pm$21.7 & 0.02$\pm$0.58 & 24.9$\pm$10.5 & 40.6$\pm$3.1 & 3.13 & 0.10 & 0.49, 0.49, 0.48, 0.52 & 9.1$^{+0.3}_{-0.3}$ & 25.16$\pm$0.12 & 7.0 & DE\\
F42\_cand00062 & & 16.3$\pm$12.0 & 0.07$\pm$0.93 & 32.8$\pm$34.0 & 15.2$\pm$4.9 & 8.08 & 0.27 & 0.75, 0.75, 0.76, 0.75 & 13.3$^{+1.8}_{-1.3}$ & 24.95$\pm$0.11 & 1.1 & $a<30$\\
F44\_cand00001 & & 40.2$\pm$21.0 & 0.03$\pm$0.57 & 15.3$\pm$0.7 & 39.1$\pm$2.6 & 3.15 & 0.10 & 0.52, 0.53, 0.50, 0.50 & 9.5$^{+0.3}_{-0.3}$ & 25.39$\pm$0.13 & 5.9 & DE\\
F44\_cand00010 & & 38.8$\pm$20.3 & 0.03$\pm$0.57 & 15.3$\pm$0.4 & 37.8$\pm$2.6 & 3.19 & 0.11 & 0.58, 0.51, 0.52, 0.54 & 8.4$^{+0.3}_{-0.3}$ & 24.16$\pm$0.05 & 9.7 & DE\\
F44\_cand00022 & & 43.7$\pm$22.7 & 0.02$\pm$0.57 & 14.4$\pm$0.9 & 42.7$\pm$2.7 & 2.85 & 0.10 & 0.53, 0.52, 0.50, 0.50 & 8.2$^{+0.3}_{-0.3}$ & 24.47$\pm$0.06 & 10.6 & DE\\
F44\_cand00030 & & 49.4$\pm$25.5 & 0.02$\pm$0.56 & 13.5$\pm$0.1 & 48.4$\pm$2.7 & 2.55 & 0.08 & 0.79, 0.78, 0.77, 0.74 & 5.7$^{+0.3}_{-0.2}$ & 22.54$\pm$0.02 & 33.1 & DE\\
F46\_cand00007 & & 37.8$\pm$19.8 & 0.02$\pm$0.58 & 19.9$\pm$0.6 & 37.0$\pm$2.6 & 3.27 & 0.11 & 0.46, 0.47, 0.46, 0.48 & 9.9$^{+0.3}_{-0.3}$ & 25.56$\pm$0.14 & 4.8 & DE\\
F46\_cand00017 & & 50.4$\pm$25.9 & 0.02$\pm$0.56 & 24.4$\pm$2.1 & 49.6$\pm$2.8 & 2.47 & 0.08 & 0.50, 0.50, 0.49, 0.49 & 8.4$^{+0.3}_{-0.2}$ & 25.33$\pm$0.13 & 9.6 & DE\\
F46\_cand00019 & & 35.6$\pm$18.7 & 0.02$\pm$0.58 & 20.2$\pm$0.6 & 34.8$\pm$2.6 & 3.45 & 0.11 & 0.54, 0.55, 0.55, 0.57 & 9.3$^{+0.3}_{-0.3}$ & 24.64$\pm$0.06 & 6.5 & DE\\
F46\_cand00024 & & 41.3$\pm$21.4 & 0.02$\pm$0.57 & 20.3$\pm$0.3 & 40.4$\pm$2.6 & 2.99 & 0.10 & 0.51, 0.54, 0.54, 0.52 & 8.7$^{+0.3}_{-0.3}$ & 24.76$\pm$0.08 & 8.3 & DE\\
F48\_cand00002 & & 45.5$\pm$22.9 & 0.01$\pm$0.56 & 61.5$\pm$0.2 & 45.0$\pm$2.6 & 3.68 & 0.12 & 0.74, 0.62, 0.71, 0.79 & 8.1$^{+0.3}_{-0.2}$ & 24.54$\pm$0.06 & 11.3 & DE\\
F49\_cand00001 & & 45.8$\pm$23.0 & 0.01$\pm$0.57 & 72.3$\pm$4.5 & 45.5$\pm$3.1 & 3.82 & 0.13 & 0.74, 0.63, 0.66, 0.76 & 8.8$^{+0.3}_{-0.3}$ & 25.32$\pm$0.12 & 8.1 & DE\\
F54\_cand00001 & & 38.3$\pm$20.1 & 0.03$\pm$0.59 & 32.0$\pm$11.4 & 37.4$\pm$3.4 & 3.28 & 0.11 & 0.65, 0.76, 0.49, 0.78 & 9.3$^{+0.4}_{-0.4}$ & 24.97$\pm$0.16 & 6.5 & DE\\
F54\_cand00002 & & 41.8$\pm$21.7 & 0.02$\pm$0.57 & 22.0$\pm$1.1 & 40.9$\pm$2.7 & 2.94 & 0.10 & 0.62, 0.72, 0.52, 0.76 & 7.3$^{+0.3}_{-0.3}$ & 23.40$\pm$0.04 & 15.9 & DE\\
F54\_cand00013 & & 36.7$\pm$19.3 & 0.03$\pm$0.58 & 23.9$\pm$3.3 & 35.8$\pm$2.8 & 3.34 & 0.11 & 0.61, 0.69, 0.55, 0.74 & 7.9$^{+0.4}_{-0.3}$ & 23.35$\pm$0.03 & 12.6 & DE\\
F54\_cand00014 & & 34.8$\pm$18.3 & 0.03$\pm$0.58 & 21.2$\pm$0.2 & 33.9$\pm$2.5 & 3.49 & 0.12 & 0.61, 0.70, 0.61, 0.82 & 10.0$^{+0.3}_{-0.3}$ & 25.20$\pm$0.17 & 4.8 & DE\\
F60\_cand00001 & & 37.6$\pm$1.2 & 0.78$\pm$0.04 & 77.1$\pm$7.2 & 8.4$\pm$1.5 & 18.40 & 0.61 & 0.76, 0.64, 0.73, 0.91 & 15.8$^{+0.9}_{-0.8}$ & 24.78$\pm$0.07 & 1.6 & DE\\
\hline
\end{tabular}
\end{center}
\footnotesize{The TNOs discovered in the Latitude Density Search with their search names (field number plus candidate number) and their designations from the Minor Planet Center (MPC).  The semi-major axis ($a$), eccentricity ($e$), inclination ($i$), distance ($d$), and absolute magnitude in the $r$-band $H_r$ are calculated from the two night arc.  The rate of motion is averaged between the two nights, and the trailing in a single exposure is calculated using the rate over the 120~s exposure. The Full Width Half Maximum (FWHM) of each image is given in arcseconds and the pixel scale of HSC is 0.17 $\arcsec$ pixel$^{-1}$; for a few images we were unable to construct a good PSF indicated by ``--'', in this case binarity was assessed on the other images only.  The instantaneous Hill radius is calculated from the $H_r$ magnitude, see text for details.  The last column indicates likely classifications; `DE' refers to `Dynamically Excited'.  For complete orbital parameters and details on how these are calculated, see \citet{chen2020}.}
\bigskip
\bigskip
\end{table}

The orbital parameters and their associated uncertainties are calculated using orbital fitting methods based on \citet{bernstein}, and the apparent $r$-band magnitudes are converted to absolute magnitude $H_r$ assuming the objects are observed at opposition (targets were 8-18 days from opposition).
The discoveries were not tracked beyond the initial 2 nights.
As a result, (except for two previously known objects identified using MPChecker\footnote{\url{https://www.minorplanetcenter.net/cgi-bin/checkmp.cgi}} provided by the Minor Planet Center) the orbital fits have significant uncertainties and the objects have been `lost'.
The orbital fits do have large uncertainty (detailed in Table \ref{tnolist}), but the distance, inclination, and $H_r$ are well measured.
The mean uncertainty is 9\% (or 0.3) on $H_r$ and 1.7$^{\circ}$ on $i$, and individual values are provided in Table \ref{tnolist}. 
The semi-major axis ($a$) and distance ($d$) from the fits, with associated uncertainties, are shown in Figure \ref{fig_dai}, which demonstrates the significant difference in the uncertainty between $a$ and $d$.

Based on the clear separation in $aei$ parameter space, it is possible to separate the sample into likely cold classical TNOs and dynamically excited TNOs.
We identified 10 of the TNOs as likely cold classicals based on their best fit orbital parameters; these objects have low inclinations $i\le3.7^{\circ}$, low eccentricities $e\le0.03$, and semi-major axes in the classical region $41.6<a<47.2$.
The objects we classify as cold classicals were well isolated in $i$ from the rest of the sample (the next lowest inclination is 7.8$^{\circ}$), and all of these had low eccentricities $e\le0.03$ (though $e$ is poorly constrained for all fits).
There are no low-$i$ objects found outside this semi-major axis range, which covers the classical Kuiper belt region.
The likely cold classicals are primarily $8\le H_r \le9$, while the larger sample of dynamically excited objects spans a wider range $5\le H_r \le 10$.

To provide context for the interpretation of our binary search, we calculated the Hill sphere of each of the detections.
The Latitude Density Search sample includes a large range of distances and absolute magnitudes, as well as a mix of cold classical and dynamically excited objects.
We converted $H_r$ to diameter for each object, assuming a typical albedo of 8.5\% (for dynamically excited objects and $a<30$ objects) or 14\% (for the likely cold classical objects), based on typical albedos from \citet{vilenius2018}.
We converted the diameter to a mass, assuming a density of 500 kg m$^{-3}$, which has been assumed to be the density of Arrokoth in previous works based on the minimum density of 290 kg m$^{-3}$ and the density of comets such as 67P/Churyumov-Gerasimenko \citep{spencer2020}.
This density is also consistent with the bulk densities presented in \citet{noll2020} for objects with masses from 10$^{18}\sim$10$^{20}$~kg, though the region between 10$^{19}$--10$^{20}$~kg and $\lesssim$5$\times$10$^{17}$~kg is not well sampled.
Our sample ranges from 10$^{14}\sim$10$^{19}$~kg, with the majority of the objects estimated to be $\sim$10$^{17}$~kg.
We calculated the Hill radius based on these mass values, $R_H$, as in \citet{noll2008ssbn}, assuming all of the mass calculated from $H_r$ is in the primary.
(In the worst case scenario for this assumption, an equal mass binary, the Hill radius would be reduced by 20\%.)
We report the Hill radius for the sample in Table \ref{tnolist} in arc seconds on sky at the time of the observation.
We have provided an instantaneous value and not an average value based on the full orbit, which \citet{noll2008ssbn} noted can be significantly different.
The Hill radius is intended to provide a metric for assessing binary detectability, which incorporates both the absolute magnitude $H_r$ of the objects and the albedo differences between dynamically excited TNOs and cold classicals.
The cold classical Hill radius values have a median of 7.8$\arcsec$ and a standard deviation of 3.5$\arcsec$, and for the brightest cold classical is 18.9$\arcsec$.
The dynamically excited objects have larger hill radii, with a median of 8.5$\arcsec$ and a standard deviation of 7.1$\arcsec$. 
The median Hill radius of the dynamically excited objects is $\sim$10\% larger and has a larger standard deviation, and thus many of these targets are more deeply probed than the cold classicals.
TNO binaries are found deep within the Hill sphere \citep{noll2004, kern2006}, and the provided Hill radius value is thus intended as a rough guide to the reader to aid in understanding the relative resolution of the target list.

In spite of the limitations in the Latitude Density Search, this sample provides a valuable snapshot of the distribution, brightness, and characteristics of TNOs with a variety of ecliptic latitudes. 
In this work we utilize the four discovery images to search for binary TNOs.
The data were acquired in photometric skies and exquisite seeing conditions of 0.4\arcsec, so they provide a useful opportunity to investigate the effectiveness of binary search methods and identify binary TNOs.

\section{Identifying Binary TNOs}
\label{sec:binary_id}

%The Latitude Density Search detection of 54 TNOs with a range of orbital parameters and consistent observing conditions provides a valuable sample for determining the relative fraction of binary objects in different populations and at different absolute magnitudes $6.4\le H_r \le10.0$.
None of the objects in the Latitude Density Search were obvious visual binaries, even with the 0.4\arcsec seeing.
In order to identify binaries and to quantify our detection limits, we subtracted a trailed Point Spread Function (PSF) using a method similar to \citet{marsset2020}.
This Trailed PSF (or TSF) subtraction analysis is described below.

\subsection{Trailed Point Spread Function Subtraction Method}

The TNOs were all discovered in sidereally tracked images, so the objects are slightly trailed in each frame.
This is a challenge for binary identification, as the TNO can appear slightly non-round in the direction of motion.
The amount of trailing for each object is included in Table \ref{tnolist}, and is typically of order 0.5 pixels or 0.09$\arcsec$ per image.
The amount of trailing is known from the rate of motion measured across the two nights, and the large number of stellar sources provide a means to robustly measure the PSF.
We calculate the rate and angle of motion from the Source Extractor astrometric positions of the TNO centroid measured as part of the discovery survey.
Even though the long-term orbit solutions are not well constrained, the rate of motion of the object between different frames is measured with high accuracy.
(For the two objects with known orbits, the rate of motion was consistent to the JPL ephemeris rate of motion to within 0.2--0.3$\arcsec$ hr$^{-1}$, $\leq0.06$ pixels per image, and using the known orbit in our analysis did not produce different results.)
For each image, we selected the sources on the chip with stellar profiles and created a model PSF.
This model PSF is a combination of a Moffat profile, calculated by minimizing the $\chi^2$ residual between the stellar profiles and the modeled Moffat profile, and a model of the average residuals between the stellar sources and the Moffat profile.
We used TRIPPy \citep{fraser2016phot} to create a model of the TSF in each image by convolving the modeled PSF of the stars with the measured rate and angle of motion of each TNO.
The position and flux of the TSF were fitted to each TNO utilizing the Python emcee implementation of the Markov Chain Monte Carlo (MCMC) fitting \citep{foreman2013}.
This method uses multiple walkers to explore the three free parameters of position ($x$ and $y$) and flux of the TNO, and determines the minimum $\chi^2$ residual between the TSF model and the image.
The single TSF subtraction utilized 20-50 walkers, 25 steps, and 25 burns.
For many of our sources, 20 walkers were sufficient to find an appropriate fit.
For some of the objects, however, 20 walkers did not provide sufficient sampling to converge to a good solution (as discussed further in the following section), so the fitting was repeated with 50 walkers in order to produce a clean subtraction.

We calculated and subtracted TSFs for all four images of each of our 60 objects.
If an object is binary, subtracting a single TSF can result in two recognisable patterns, depending on the relative brightness and separation.
When the secondary is fainter than the primary, then the fitter selects a position close to the center of the primary, often slightly over-estimating the primary brightness, and the secondary is exposed in the residuals next to a slight over-subtraction.
When the two components are of similar size or magnitude, the fitter typically places the TSF between the two peaks and produces a butterfly pattern; see \citet{marsset2020} and Section \ref{sec:sensitivity} for examples.
Once the TSF was subtracted, we searched for these signatures in the residuals.

\subsection{Searching the Residuals for Binaries}
\label{sec:binary_resid}

Identifying the binary TNOs required a careful examination of the residuals after the TSF subtraction to determine whether there is sufficient un-subtracted signal or a significant over-subtraction in the residuals which could indicate a binary source.
We used two identification methods for finding significant residuals: a manual visual check and a numerical estimate of the residual flux.
For the manual visual check, we conducted three independent inspections of all image residuals searching for the over-subtractions and under-subtractions typical of binary TNOs.
We compiled lists of objects of interest based on this manual inspection.

We also developed a metric for quantifying the residuals relative to the background level.
To ensure that we developed an appropriate metric for identifying binary residuals, we first implanted binary sources into a clear background region in a representative image.
We generated double TSFs using TRIPPy and implanted these fake binary sources with a range of separation distances, angles, and relative brightnesses into the image.
We then subtracted a single TSF fit from the implanted source to reveal the residuals.
We were able to identify the majority of the implanted binary sources by examining a 10 square pixel region around the TNO center.
We selected only those pixels where the flux was more than 2$\sigma$ discrepant from the median background (calculated over a sigma-clipped 200$\times$200 pixel cutout), and summed the difference between the flux in each pixel and the median background.
We created a separate sum for `positive' and `negative' discrepant values, and determined a threshold for flagging the residuals.
(The overwhelming majority of flagged subtractions have positive residual flux, and most flagged for negative flux were also flagged for positive flux.)
We also investigated using a cut depending on the number of pixels where the 2$\sigma$ threshold was reached, and the results are nearly identical to the method which uses the sum of the pixels.
We applied this criteria to the TNO discovery images after subtracting the TSF.
This method resulted in a longer list of candidates than the visual inspection.
These included: 4 cases where a background source fell within the cutout and 2 with unusual background structure within the cutout but not directly on the subtraction, 10 cases of a poor centroid calculation which was resolved by increasing the number of walkers to 50 and repeating the subtraction, 5 cases of a poor flux fit resulting in a slight over-subtraction (these only occurred in a single frame of four for each object, and were resolved with a second fit utilizing better initial conditions), and the 7 objects also identified in the visual inspection, which we followed up with additional testing.

We further analyzed the possible binaries flagged by both methods: 7 candidate binaries objects which we tested extensively (F06\_cand00002, F40\_cand00007, F40\_cand00009, F40\_cand00053, F40\_cand00057, F44\_cand00030, F54\_cand00002).
We inspected the PSFs used in the subtraction to test whether the residuals were due to a poor PSF calculation.
This included carefully examining the profile of each star that was used for the PSF model for contamination from faint nearby sources, saturated star bleeding corrections (as part of the Hyper-Suprime Cam processing), and other residuals that might contaminate the PSF.
One of the sources (F44\_cand00030) was discovered in an edge-chip of the camera, near the edge of the camera field of view, and the distortion of the PSF varied noticeably across the chip.
This distortion had resulted in the illusion of an apparent secondary.
As a result, on our second analysis of this candidate we limited our PSF stars to those within a few hundred pixels of the TNO, which produced a significantly different PSF than the full chip.
We were able to dramatically improve the subtraction, but not entirely eliminate the residuals, likely due to the small number of stars sufficiently close to the TNO.
As a result, that PSF is of poorer quality than the TNOs for which the full chip provides an excellent measure of the PSF.
A few of candidates that had bright residuals that were due to a poor centering of the TRIPPy MCMC fitter during the broader analysis, and a good single subtraction was obtained with a time intensive re-computation using more walkers (50-70) and a variety of starting $x$ and $y$ coordinates for the TNO.
Each of these steps was significantly more time intensive than the previous analysis of the entire data set, particularly the manual inspection of each star in the PSF/TSF and the increased computation time due to the additional walkers and initial conditions used in the MCMC fits.
Based on this expanded analysis, we determined that 6 of the objects were false positives which could be well subtracted by a single TSF, and one of the objects was binary, F40\_cand00053.

\subsection{Fitting the Binary Components}

\begin{figure}[h]
\begin{center}
\includegraphics[width=.7\textwidth]{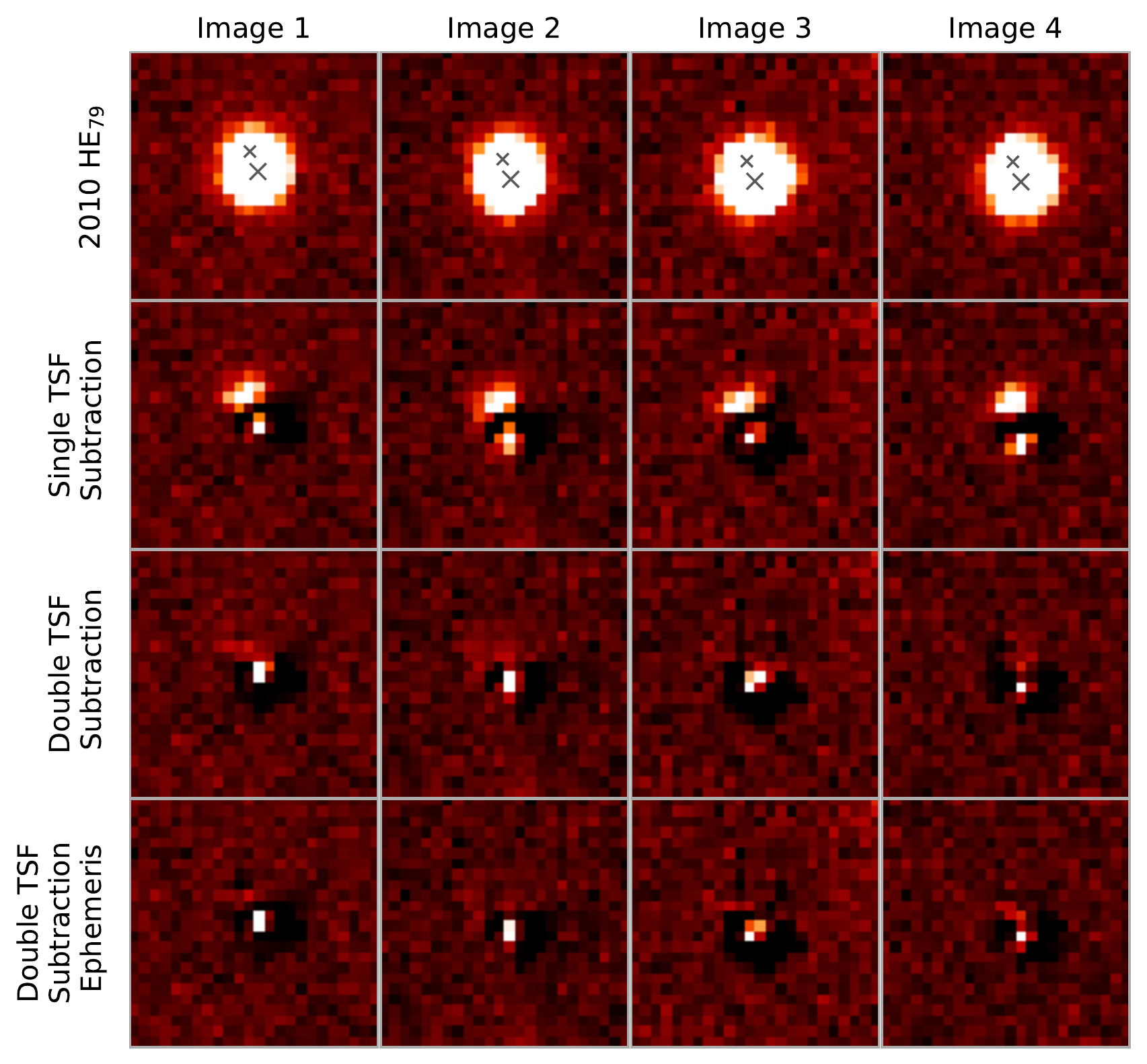}
\caption{
The \textbf{upper row} shows the unsubtracted images of 471165 (2010 HE$_{79}$); images 1 and 2 were acquired on the first night and 3 and 4 were acquired on the second night of observations.
The position of the primary and secondary are indicated by a larger and a smaller 'x'.
The \textbf{second row} shows the TNO after a single TSF was subtracted.
The four images of 471165 (2010 HE$_{79}$) show similar residuals for a single TSF subtraction.
The primary source is removed (with a slight over-subtraction due to contamination by the secondary) and the secondary is revealed.
The \textbf{third row} shows the results of subtracting two TSFs with a 2.09 pixel separation at -21.9$^{\circ}$ and a secondary brightness of 17\% the primary brightness.
The residuals are significantly improved with the removal of two TSFs, but not completely consistent with the background.
The top three rows use a rate and angle of motion calculated from the discovery images, while the \textbf{bottom row} shows the results of the subtraction when the rate and angle of motion is computed using from the JPL ephemeris service, different by 0.2$\arcsec$ hr$^{-1}$ and 0.8$^{\circ}$.  This results in a small improvement to the residuals after subtraction, but the similarity of these results demonstrates that the rate and angle determined from the discovery images is sufficient for this analysis.
}
\label{fig_binary_images}
\end{center}
\end{figure}

The TNO which was identified as a binary object in the above analysis, F40\_cand00053 or 471165 (2010 HE$_{79}$), was modeled as a two-component system.
For this analysis, we used the TRIPPy fitting function fitDoubleWithModelPSF to simultaneously fit two TSFs to the TNO.
This function utilizes the trailed PSF and an initial guess for the starting positions of the two objects and their flux ratio, and attempts to determine the preferred solution using MCMC.
We used the MCMC fitter with 80 walkers to determine an acceptable subtraction of two TSFs.
We used the rate and angle of motion determined from the four images in our survey for the initial analysis, however, we also tested the JPL Horizons ephemeris\footnote{\url{https://ssd.jpl.nasa.gov/horizons.cgi}} rate and angle of motion.
This gave nearly identical results, and we found that our calculated rate and angle of motion was extremely close to the JPL ephemeris value, only different by 0.2$\arcsec$ hr$^{-1}$ and 0.8$^{\circ}$, or 0.04 pixels during a 120 second exposure.
We did not stack the four images of the TNO because the stellar background changes significantly between the two nights; the small overlap in comparison stars means that it would be difficult to measure the PSF for the stack.
Based on the results from all four images of the TNO, we determined a single set of parameters which provided a good subtraction for the TNO in all images: separation distance, separation angle, and relative brightness, and present the results in Figure \ref{fig_binary_images}.

The confirmed binary object is obvious in the subtractions.
The single TSF subtraction centered on the primary, and revealed the remaining secondary component, see Figure \ref{fig_binary_images}.
The double TSF subtraction (using either the rate of motion calculated from the images or extracted from the JPL ephemeris) produced an acceptable fit for all four images of the TNO, which included the results of the fit (separation, brightness ratio) which were similar for all four individual images.
The consistency of the brightness ratio and separation between the primary and secondary across all 4 images, acquired in 4 separate visits spread over 2 nights, with varying airmass and slightly different delivered image quality, is compelling evidence that this object is binary. 
We calculated average offset and brightness ratio, and thus determined the brightness ratio and pixel separation of the primary and secondary in $x$ and $y$ and the resulting distance.
This average was examined for all four images, and minor iterative changes were made to the weighting of each solution to produce the best residuals overall.
The binary TNO 471165 (2010 HE$_{79}$) has combined brightness of $m_r$=20.88$\pm$0.00 a separation of 2.09 pixels or 0.36$\arcsec$ and the secondary is 17\% the brightness of the primary ($\Delta m$=2.0 magnitudes).
This binary was easily detectable in the residuals and well above our detection limits.

\section{Sensitivity to Binary TNOs}
\label{sec:sensitivity}

\begin{figure}[h]
\begin{center}
\includegraphics[width=1.\textwidth] {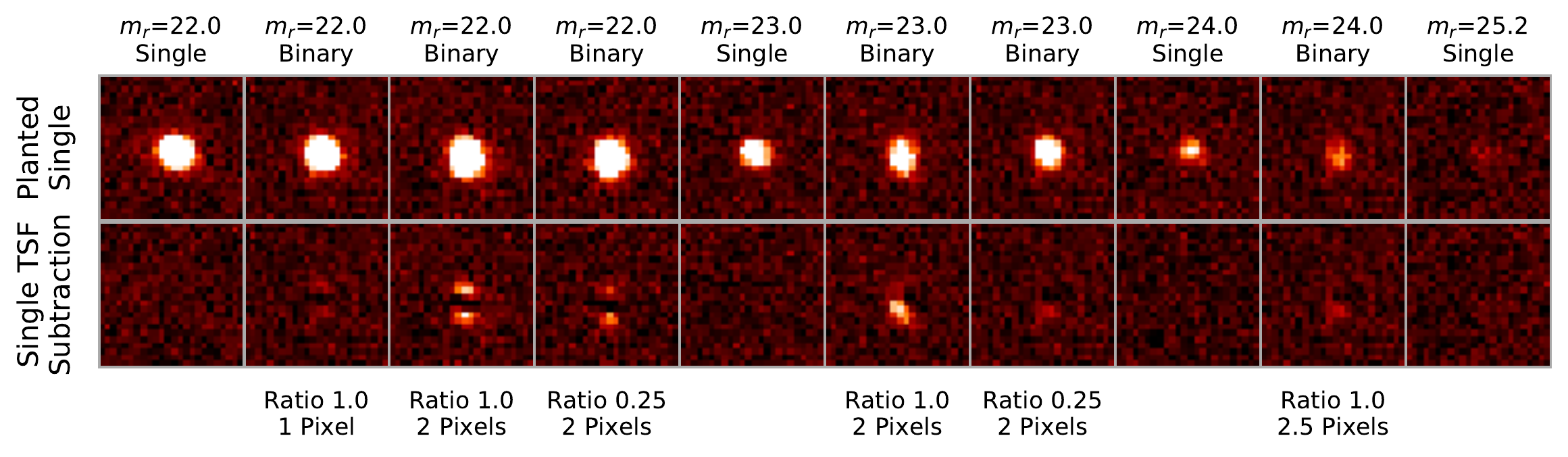}
\caption{The planted single and binary sources before and after TSF subtraction.  The \textbf{upper row} shows the planted sources.  This includes planted Single TSFs as well as binary TSFs, as indicated for each column.  The secondary source is planted directly below the primary along the y-axis.  The brightness ratio and pixel separation of the two planted sources are given below the binary images.  The \textbf{bottom row} shows the results of a single TSF subtraction.  The residuals for the single source subtractions are consistent with the background, while the single subtraction of a TSF from the binary sources have significant residuals. These typically show one of two patterns: the fitter centers the TSF in the middle and leaves an over-subtraction with bright wings, or the fitter selects a position between the two object centers where the flux of the sources overlap. This results in an over-subtraction next to a residual bright spot. All planted sources (single and binary) are moving at 3.0$\arcsec$ hr$^{-1}$.
}
\label{fig_planted}
\end{center}
\end{figure}

We tested the effects of different characteristics of binary pairs on their detectability by implanting binary and single sources into our data.
We selected an image which was representative of the seeing conditions for at least half of the images of each target.
The seeing conditions were slightly better on first night of observation than the second. 
We chose an image with FWHM of 3.02 pixels (0.51$\arcsec$) and selected 40 positions on the image where there were no background sources within 30 pixels to contaminate our implanted objects.
We implanted two TSFs with a typical rate of motion for the objects in our sample (3.0$\arcsec$ hour$^{-1}$) and a range of combined magnitudes (22.0-25.5 in steps of 0.1 magnitudes), separations (1.0--2.5 in steps of 0.5 pixels), and relative brightness ratios (1.0, 0.5, 0.25).
We also implanted single objects with the same range of magnitudes for comparison.
For each parameter combination we planted the binary or single object in $\ge$30 locations on the image.
We used the same processing script as for our real detections to subtract a single TSF from the implanted sources, however, we increased the number of walkers to 50 which was needed to verify our candidate binaries.
This resulted in a range of residuals depending on the input parameters, shown in Figure \ref{fig_planted}; some identifiable as binary objects and some indistinguishable from the single objects.

Using the planted binary and single sources, we developed the automated identification of residuals which resembled binary objects, described \ref{sec:binary_resid}.
For the automated search, we calculated the sum of the flux minus the median for the pixels which were 2$\sigma$ greater than the median background flux.
We determined the threshold for the positive and negative summed flux which was 2$\sigma$ discrepant from the mean which we used for identifying possible binary sources.
We used visual spot checks to confirm the validity of this criteria.
We also tested which input parameters affected the likelihood of binary identification for this data.
During our initial testing isolating different parameters, we determined that varying the angle of orientation relative to the direction of motion to 0$^{\circ}$, 45$^{\circ}$ and 90$^{\circ}$ had no measurable impact on our ability to identify the object as binary, likely because the amount of motion in each 120 second image was small.
Our initial tests also included separations of up to 5 pixels, and we determined that at separations $\ge$2.5 pixels, binary objects were reliably detected for any secondaries brighter than the survey limit, while at 2 pixels or less of separation, limits were brighter than the survey limit. 
For the fake planted sources at known positions, we were able to remove false positives due to poor centroiding in the fit.
For the real objects with poor centroiding, we had repeated the fitting process with additional walkers and/or starting positions, but for the planted sources with 1-2.5 pixels of separation, we simply excluded from the results any fit where the distance from the planted primary was $>3$ pixels.
Based on this analysis, we present a range of detectability limits calculated based on the relative brightness and separation distance between the binary pair.

\begin{figure}[h]
\begin{center}
\includegraphics[width=1.\textwidth]{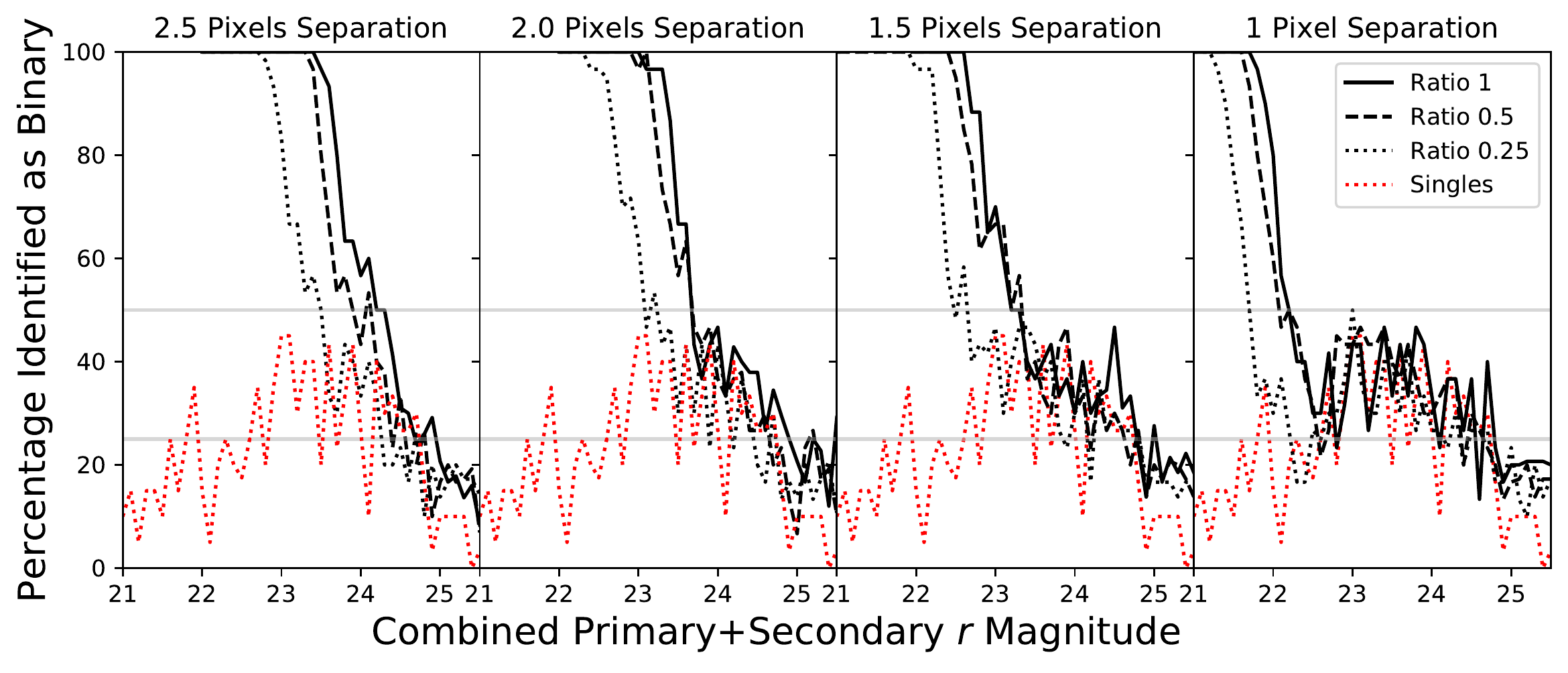}
\caption{
The percentage of planted sources identified as binary.  We implanted binary and single TSFs at a range of $r$-magnitudes, separation distances, and relative brightness ratios.  For each set of conditions, we planted $\ge$30 sources, subtracted a single TSF, then evaluated the residuals for evidence of binarity.  We planted `single' (red dotted) TSFs as a measure of our false positive rate. (Note that these do not have a `separation', so the same data is plotted for the singles in all panels.)  We used a 50\% detection threshold to determine our limiting magnitude presented in Table \ref{maglimtable}.
Binary 471165 (2010 HE$_{79})$ is $m_r$=20.88$\pm$0.00, in the 100\% detection efficiency region of the plot.}
\label{fig_detectionrate}
\end{center}
\end{figure}

\begin{figure}
\begin{center}
\includegraphics[width=0.9\textwidth] {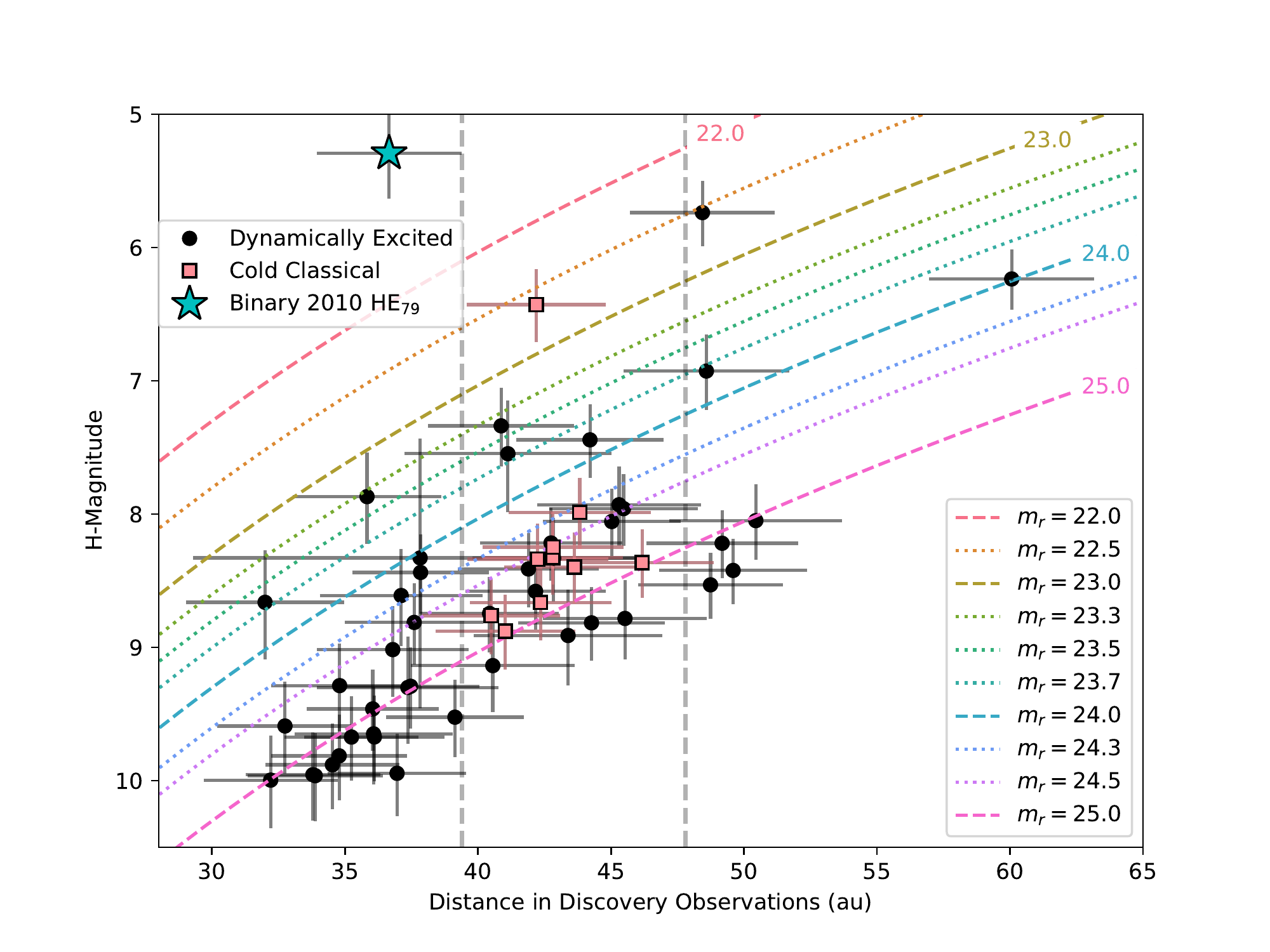}
\caption{
The Latitude Density Search detections and $H_r$-magnitudes with dotted/dashed lines of constant magnitude. The distance at discovery of the objects and their absolute $H_r$-magnitudes are shown.  (Five objects with distances 12--28 au in Table \ref{tnolist} are omitted from the figure.) The dashed gray vertical lines indicate the locations of the 3:2 and 2:1 resonances. The cyan star is the binary TNO 471165 (2010 HE$_{79}$), the pink squares are non-binary likely cold classical TNOs, and the black circles are non-binary dynamically excited TNOs.  The dotted lines indicate the $r$-band magnitude of objects as a function of the distance and $H$-magnitude, as indicated: $m_r$=22.0, 22.5, 23.0, 23.3, 23.5, 23.7, 24.0, 24.3, 24.5, 25.0.  The $m_r$ detection limits for consistent binary detection, based on the intrinsic separation and brightness ratio different binary configurations, are listed in Table \ref{maglimtable}.  The identified binary detection has a 2 pixel separation (0.34$\arcsec$) and a brightness ratio of 17\%.  For the separation of two pixels and a brightness ratio $\ge$50\%, 7 objects in the sample are above the binary detection limit $m_r\le23.7$.
}
\label{fig_limits}
\end{center}
\end{figure}

The results of our sensitivity limit testing are presented in Figure \ref{fig_detectionrate}.
The single planted objects (red dotted line on Figure \ref{fig_detectionrate}) show the rate of false positive identification of binaries.
This false positive rate is quite low for a combined $m_r\lesssim22$, and slightly higher near the detection limit for binary identification.
Based on this false positive rate and the detection fraction for the planted binaries, we have selected our detection limit as the magnitude where 50\% of the planted sources identified as binary.
Because we did extensive investigation into any sources where at least two of the four single TSF subtractions showed evidence of possible binarity, this is a reasonable estimate of the detection limit for binaries in this search.
The planted binaries also showed significant correlations between the residuals for the series of images, which are easily noted in a visual inspection.
These detection limits are provided in Table \ref{maglimtable}, and the survey discoveries with lines of constant magnitude are shown in Figure \ref{fig_limits}.

The careful computation of the binary detection limits provides a critical tool for assessing the sensitivity of our data to binary TNOs.
At separations $\ge2.5$ pixels, or 0.4$\arcsec$, secondary sources brighter than the Latitude Density Search detection limit $m_r\simeq25.2$ were identifiable.
In Table \ref{maglimtable}, we provide the combined magnitude of the primary and secondary as well as the magnitude of each.
In general, this detection method is more sensitive to brighter secondaries, however, the fitting of the TSF to the source leaves the most noticeable residuals if the primary dominates the signal, so it is also more sensitive to bright primaries.
Because we are measuring our sensitivity as a function of the combined brightness, as we increase the brightness of the secondary at a single total magnitude, we decrease the brightness of the primary.
This tension is best seen visually in Figure \ref{fig_detectionrate}, where the detection efficiency with respect to the combined brightness for equal mass binaries and binaries with a brightness ratio of 0.5 are nearly identical.
Where both the primary and secondary are well above the detection limits (e.g. $m_r$=21), we detect binaries at near 100\% efficiency even with a 1 pixel separation.

The binary detected in our sample, 471165 (2010 HE$_{79}$), is one of the brightest and the smallest $H$-magnitude in our detection list, and we expect a detection efficiency of $\sim$100\% of the binaries with this apparent separation and magnitude ratio, with a $\sim$10\% false positive rate.
The binary object is significantly brighter than our detection limit, and binaries with a similar separation and companions with a brightness ratio $\ge$0.5 could have been detected for an additional 6 TNOs, including the brightest cold classical in the sample.
This is a binary fraction of $\sim$17\% for the dynamically excited objects in our sensitivity limits.

\begin{table}[h!]
\setlength{\tabcolsep}{2.5pt}
\caption{High Latitude Search Discoveries}
\label{maglimtable}
\begin{center}
\begin{tabular}{ c  c  c  c  c  c  c c c c} 
Separation & Brightness Ratio & Magnitude Limit & Primary & Secondary\\
(Pixels) & (Primary/Secondary) & (Combined $m_r$) & $m_r$ at limit & $m_r$ at limit \\ \hline
2.5 & 1.0 & 24.3 & 25.1 & 25.1 \\
 & 0.5 & 23.9 & 24.3 & 25.1 \\
 & 0.25 & 23.5 & 23.7 & 25.2 \\\hline 
2.0 & 1.0 & 23.7 & 24.5 & 24.5 \\ 
 & 0.5 & 23.7 & 24.1 & 24.9 \\ 
 & 0.25 & 23.2 & 23.4 & 24.9 \\ \hline 
1.5 & 1.0 & 23.3 & 24.1 & 24.1  \\ 
 & 0.5 & 23.3 & 23.7 & 24.5 \\  
 & 0.25 & 22.7 & 22.9 & 24.4 \\ \hline 
1.0 & 1.0 & 22.2 & 23.0 & 23.0 \\ 
 & 0.5 & 22.1 & 22.5 & 23.3  \\  
 & 0.25 & 21.7 & 21.9 & 23.4  \\ 
\hline
\end{tabular}
\end{center}
\footnotesize{The magnitude limit is computed based on the combined brightness of the primary and secondary objects, as this provides the most direct comparison to the real objects.
At separations of 2.5 pixels, the magnitude limits are consistent with detecting secondaries that are brighter than the survey detection limit, $m_r\simeq25.2$.
At larger separations, we expect this same limiting magnitude, and that the objects would begin to appear visually binary.
The magnitude limit is the magnitude where at least 50\% of the planted objects were flagged as binary in the automated residual check, interpreted as the limit at which we would have identified the binary based on our 4 images with our automated search.}
\bigskip
\bigskip
\end{table}

\vspace{0.5cm}

\section{Discussion}
\label{sec:discussion}

The excellent seeing conditions of the observations from the Latitude Density Search and the 54 TNO detections provided a diverse sample for a binary TNO search and a useful data set for a thorough characterization of binary detection efficiency in near-optimal ground-based conditions.
This sample included many faint detections, with a ($5\sigma$) limiting discovery magnitude of $m_r\simeq25.2$.
We were not able to increase our signal to noise by stacking the four images of each object due to the poor quality of the orbit fits and differences in the background stars available between the two nights, so we tested each of the 4 images of each target for binarity.
If stacked images are poorly aligned, they could appear to be multiple TSFs, and imitate a binary object.
If the objects had been tracked and their orbits had been well-determined, however, we could have verified the accuracy of their positions and rates of motion, and stacked the images in each night.
We still would have had difficulty stacking across two nights if there were not sufficient overlapping background stars.
Stacking the four images we acquired would have increased the signal to noise of approximately half of the sample (both the cold classicals and the dynamically excited TNOs) sufficiently for the detection of an equal brightness binary pair with a 2 pixel separation.

Based on this analysis, we recommend this method of multiple TSF fitting for the detection and tracking of binary TNOs, either bright objects in single images or stacked images of fainter known TNOs acquired in a single night.
Tracking at the sidereal rate provides a good measure of the stellar PSF and accurate astrometry, and for ground-based analyses we expect better results from tracking sidereally and stacking images if a long exposure time is needed to achieve the required signal to noise.
An application of this search method and the sensitivity limit analysis to anticipated data from the Rubin Observatory Legacy Survey of Space and Time (LSST) could result in the detection of a significant number of new binary TNOs and an understanding of their intrinsic distribution.
Additionally, the two TSFs provide a astrometric positions for the components of the binary, and this could be an effective way to do long-term tracking of known binary TNO orbits.
Fitting multiple TSFs could also be used to extract the separate positions of known binary TNOs from archival data to provide constraints on fitting the orbits of the binary pair at past epochs.
We hope that the detection limits and methods described in this work will prove useful for identifying archival data and specific targets which would be appropriate for such analyses.
Tracking known binaries with this method has additional advantages, for example, if characteristics of the binary are already known (such as the brightness ratio) from other facilities, these can be fixed in the fitting process to reduce computation time.
Fitting multiple TSFs can also be used to extract multi-band photometry of binaries which are not visually resolved, as in \citet{marsset2020}.
Based on our analysis, we expect that ground-based data acquired in optimal seeing conditions can be used to complement space-based observations of binary TNOs.

A direct comparison of binary detection rates in the various available binary searches is complicated by observing biases-- the detection limits depend on the brightness and apparent separation of the binary pair.
Previous works have attempted to measure the intrinsic fraction of binary TNOs.
\citet{noll2003} conducted a binary search, HST Deep, which had resolution of 0.15$\arcsec$ \citep{noll2003, kernelliot2006}, comparable to 1 pixel separation for the Latitude Density Search.
This binary search included a mix of dynamical classifications, and had a binary occurrence rate of 5\%$\pm$2\% \citep{noll2003}.
The Magellan binary search, also targeting a mix of dynamical classes, had a resolution of $\ge$0.3$\arcsec$ and a binary detection rate of 1.4\%$_{-0.4}^{+1.3}$ \citep{kern2006, kernelliot2006}, more comparable to a 2 pixel separation in this search.
For the dynamically excited objects above our sensitivity limits (2 pixel separation and $\ge$0.5 brightness ratio), we have a binary fraction of 17\%, but our binary fraction for the complete dynamically excited sample is 2\%.
This is most directly comparable to the Magellan binary search which quotes the 1.4\% limit as being for secondaries within 1 magnitude of the primary, or a ratio of $\sim$0.4, described as not bias corrected and calculated by simply dividing the number of binaries by the number of objects observed \citep{kernelliot2006}.
The directly comparable value for this work would be 1 of 56$\simeq$1.8\% (with large uncertainty given the single detection), which is clearly compatible with the published ratio from \citep{kernelliot2006}.
However, the bias corrected value of 17\% that we calculate is more reflective of the intrinsic binary fraction of dynamically excited TNOs within our sensitivity limits.

The identification of the brightest target as the only binary is a result similar to other discovery surveys not specifically targeting binaries.
\citet{sheppardragozzine2012} conducted multiple discovery surveys using Subaru and Magellan, and discovered $\sim$1000 TNOs in seeing conditions of 0.45-0.8$\arcsec$.
Their visual binary search only identified a single binary pair, the brightest and one of the largest in the sample, in the 3:2 resonance.
The measured wide binary fraction was inconsistent with previous brighter discovery surveys, and was interpreted as a reduced rate of occurrence of faint wide binaries \citep{sheppardragozzine2012}.
Their visual binary search is affected by similar biases to the Latitude Density Search, affecting the projected size of the Hill sphere, but they note the existence of several known binaries within their detection limits \citep{sheppardragozzine2012}.
The TSF subtraction binary search method used here is sensitive to closer separations than a visual inspection, and has also been used to search for binaries in the Colours of the Outer Solar System Origins Survey (Col-OSSOS) data \citep{marsset2020}.
Three binaries were identified in a TSF search, and all three of the Col-OSSOS binaries are $6.4\le H_r \le 6.6$, while most of the survey targets were $H_r>7$ \citep{fraser2017,marsset2020}.
The seeing conditions were less uniform for the Col-OSSOS survey than for this work, but the data were acquired with a better pixel scale (0.08$\arcsec$ pixel$^{-1}$) and signal to noise of $\sim30$, so a detailed analysis to quantify the sensitivity of their TSF analysis to binaries would provide an interesting comparison to this work.

We detected one binary object in the Latitude Density Search data and quantified the $m_r$ detection limit for binary TNOs in this search depending on the brightness ratio and projected on-sky separation of the binary pair.
The orientation angle of the pair relative to the range of motion was found to be unimportant for detectability binary TNOs when a TSF is used in the analysis.
The apparent motion of the TNOs also does not significantly impact our ability to identify binaries, as this motion is accurately measured between the images or by using the ephemeris (for objects where the orbit is known).
The small effects of TNO motion relative to the stars during the 120 second exposures is removed by using a TSF analysis.
The detection limits (for the combined brightness of the primary plus the secondary) are presented in terms of pixel separations and brightness ratio of the pair.
For magnitudes brighter than $m_r\lesssim22$, the search was sensitive to binaries with $\ge$1 pixel (0.17$\arcsec$) separation.
We were sensitive to fainter secondaries for larger separations, and once the separation reached 2.5 pixels we were sensitive to secondaries $m_r\lesssim$25.2, the survey limiting magnitude.
For the intermediate separations, we found that the detection efficiency as a function of the total combined magnitude did not change significantly for brightness ratios 0.5--1, see Table \ref{maglimtable}, even though this changes both the magnitude of the primary and the secondary.
As the secondary brightness ratio was reduced to 0.25, the detection limit is much brighter, essentially the same secondary brightness limit as the 0.5 brightness ratio secondary.
The binary detection limits provide a useful context for interpreting the discovery of a binary TNO in our sample.

\begin{figure}[h!]
\begin{center}
\includegraphics[width=0.9\textwidth] {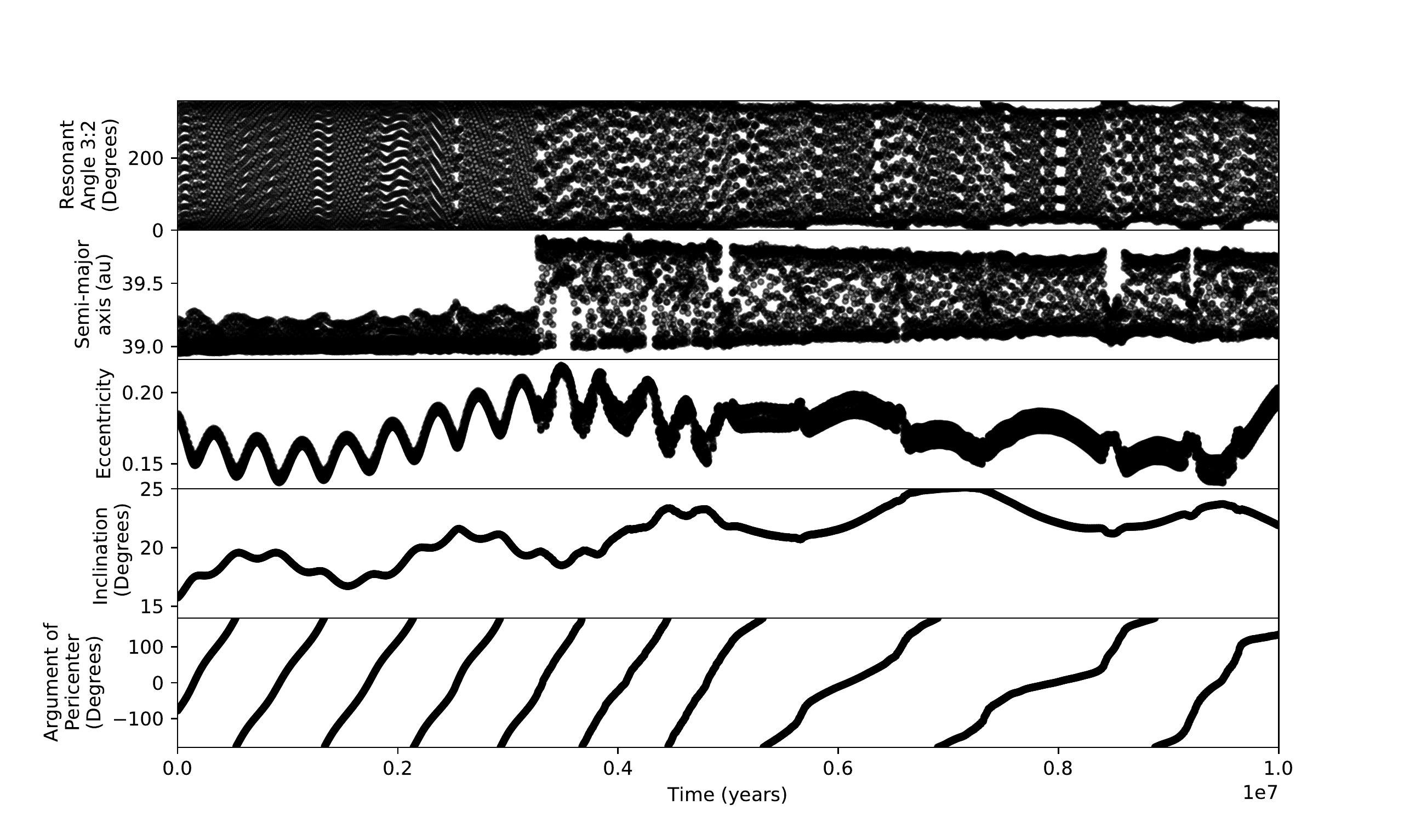}
\caption{
The binary TNO 471165 (2010 HE$_{79}$) integrated forward 10~Myr using REBOUND.
The resonant angle $\phi$ circulates for the first 4~Myr, which indicates that the object is not currently resonant.
However, the object enters the resonance with a large libration amplitude several times in the following 6~Myr.
The argument of pericenter $\omega$ does not oscillate; the binary object is not in Kozai resonance.
The object has a semi-major axis just sunward of the stable phase space of the 3:2 resonance.
}
\label{rebound}
\end{center}
\end{figure}

The binary object is the largest TNO in our sample, a previous discovery with a 28 year arc, 471165 (2010 HE$_{79}$).
The orbital elements reported in Table \ref{tnolist} are based on the 2 night arc, but more accurate coordinates are available in the Minor Planet Center, which has an orbital fit for this object with semi-major axis $a=39.2$, eccentricity $e=0.18$, and inclination $i=15.7^{\circ}$.  
As this is near the edges of the stable region for Plutinos \citep{morbidelli1997, tiscareno2009}, we integrated the object using REBOUND \citep{rein2012}.
We integrated the Sun and 8 planets for 10~Myr using the IAS15 integrator \citep{rein2015} and a timestep of 0.016 years.
We calculated the resonant angle $\phi_{32}$ with respect to Neptune, and saw both non-resonant and resonant behaviour within the classification timescale, shown in Figure \ref{rebound}.
For the first 4~Myr, the resonant angle $\phi$ circulates, indicating non-resonant behaviour.
After that, the object enters and exits the resonance repeatedly, with large libration amplitudes.
The argument of pericenter $\omega$ does not oscillate, so this object is not in Kozai resonance at any point during the integration.
This entering and exiting resonance is common for resonance-sticking objects, which typically also exhibit scattering behaviour, often jumping between the distant resonances \citep{lykawkamukai07}.
However, scattering objects experience regular perturbations by Neptune, and this object remains in the 3:2 region.
Additionally, the high inclination ($i=15.7^{\circ}$) of this TNO is consistent with the typically high-$i$ binary Plutinos \citep{noll2020}.
The semi-major axis is just sunward of the 3:2, so a reasonable origin scenario for this TNO would be that it was captured into the 3:2 in the era of Neptune's migration and moved outward (similar to Pluto), but was not deeply captured, and dropped out of the unstable phase space at the resonance edges in the late stages of migration, as Neptune jittered or Neptune's eccentricity circularized.
The presence of resonant or near resonant binaries such as this one can provide useful insight into resonance capture during Neptune's migration \citep[e.g.][]{murray-clay2011}.

This dynamically excited binary, 471165 (2010 HE$_{79}$), and may have evolved into its current orbit via the 3:2 resonance.
Assuming the current separation, 0.355$\arcsec$ or 0.9\% of its Hill radius, is representative of the semi-major axis of the pair ($a_B$) with primary and secondary radii ($r_1$, $r_2$) converted from $H_r$ assuming 5\% albedo, this TNO has a binary binding tightness of 
\begin{math} 
\frac{a_B}{\sqrt[3]{r_1^3 + r_2^3}}\sim40 ,
\end{math}
and would be classified as a `wide binary' with a 10-50\% survival probability in the dynamically excited population \citep{nesvorny_vokrouhlicky2019}.
The dynamical survival of the Plutinos is only slightly reduced relative to the other excited populations \citep{nesvorny_vokrouhlicky2019}, so this appears to be a plausible capture scenario for this TNO.
The majority of our sample for which binarity could have been easily detected are dynamically excited objects; 6 dynamically excited objects were above the limits for a 2 pixel separation and a secondary at least 50\% the primary brightness (13 with a 2.5 pixel separation).
One of these objects was binary, or 17\% (8\%), which, given the small number of objects within our sensitivity limits, is roughly consistent with the expected binary fraction for dynamically excited objects from the literature of 3\%-10\% \citep{noll2008}.
The brightest object being the only detected binary may hint that size is a factor in binarity for dynamically excited objects, however, a larger sample is needed to disentangle observation biases (such as apparent Hill radius) from differences in the intrinsic distribution.

We did not identify any cold classical binaries in our sample, but a similar search focused on cold classical binaries could improve on these results.
The large and bright cold classical 469610 (2004 HF$_{79}$), a known object which was detected in the Latitude Density Search, was well above our detection limits and showed no evidence of binarity.
The bulk of our sample of 10 cold classicals was slightly fainter than our detection threshold of $m_r\simeq$24.3 for a 2.5 pixel separation, though these are smaller objects and a 1--2 pixel separation ($m_r\simeq$22.2--23.7) is likely more appropriate.
Based on the magnitude distribution of the cold classicals in our survey and our calculated detection limits, we can determine an effective observational strategy for a ground-based binary detection survey.
In similar conditions (seeing $\simeq0.4-0.55$ and photometric skies), observing a similar field of view using 120 second exposures for a series of 6 exposures would provide an additional 1 magnitude of stacked depth, pushing the majority of the cold classical objects above the 2 pixel detection limit.
A series of at least 6 exposures would provide better sampling of the object's motion, so even if the objects were not tracked long-term, the rate and angle of motion could be constrained sufficiently well for stacking.
All of the cold classicals in the Latitude Density Search are in a single pointing of HSC, so we would expect a survey with $\sim$10 pointings and this design to provide data sufficient to determine the binarity of $\sim100$ cold classicals $H_r\lesssim9$ in approximately 2.5 hours.
This observing plan could be easily scaled to increase the signal to noise sufficiently to probe 1-2 pixel separations.

The Latitude Density Survey detected 56 TNOs, and the largest object 471165 (2010 HE$_{79}$) was found to be the only detectable binary TNO in the sample.
Based on this detection and our detection limits, the intrinsic binary fraction for dynamically excited objects with separations of 0.34$\arcsec$ and brightness ratios of $\ge$0.5 is 17\%.
Ignoring any detection biases, the binary fraction of our sample is 2\%, comparable to previous works with similar resolutions which did not quantify the detectablity of binarity for each object in their sample.
Unfortunately, as the targets in the Latitude Density Search were not tracked, additional followup and binary searches on the non-binary objects from this sample are not possible, however, a survey designed based on our results using stacked images and TSF subtraction could make significant progress in measuring the intrinsic binary fraction of TNOs at a range of $H$-magnitudes.
A larger sample of TNOs sampling a range of positions in the belt, ideally including the dynamically excited TNOs as well as and the cold classical belt would be required to robustly determine the intrinsic distribution of binary TNOs.

\acknowledgements

The authors acknowledge the sacred nature of Maunakea, and appreciate the opportunity to observe from the mountain.  This work was based on data collected at Subaru Telescope, which is operated by the National Astronomical Observatory of Japan.  Simulations in this paper made use of the REBOUND code which is freely available at \url{http://github.com/hannorein/rebound}.  J.~K. acknowledges support from the ASIAA Summer Student Program.
The authors thank Wesley Fraser and Micha\"el Marsset for their useful discussions and insight into TRIPPy and binary TNO analysis.
% The Pan-STARRS1 Surveys (PS1) have been made possible through contributions of the Institute for Astronomy, the University of Hawaii, the Pan-STARRS Project Office, the Max-Planck Society and its participating institutes, the Max Planck Institute for Astronomy, Heidelberg, and the Max Planck Institute for Extraterrestrial Physics, Garching, The Johns Hopkins University, Durham University, the University of Edinburgh, Queen’s University Belfast, the Harvard-Smithsonian Center for Astrophysics, the Las Cumbres Observatory Global Telescope Network Incorporated, the National Central University of Taiwan, the Space Telescope Science Institute, the National Aeronautics and Space Administration under Grant No. NNX08AR22G issued through the Planetary Science Division of the NASA Science Mission Directorate, the National Science Foundation under Grant No. AST-1238877, the University of Maryland, and Eotvos Lorand University (ELTE) and the Los Alamos National Laboratory. 

\facilities{Subaru} 
\software{TRIPPy, HSC Pipeline, REBOUND}

\bibliographystyle{aasjournal}

\end{document}